\documentstyle[12pt, aasms] {article}

\begin{document}

\title{An SSPM-Based High-Speed Near-Infrared Photometer for Astronomy}

\author{S.S. Eikenberry, G.G. Fazio, S.M. Ransom}
\affil{Harvard-Smithsonian Center for Astrophysics}
\affil{60 Garden St., Cambridge, MA 02138}

\begin{abstract}

	We describe the design, operation, and performance of a new
high-speed infrared photometer using the Solid-State Photomultiplier
(SSPM) detector.  The SSPM was developed by Rockwell International
Science Center and has single-photon counting capability over the
0.4-28 micron wavelength range, intrinsic time response of order 1 ns,
and low detector noise (Petroff, {\it et al.}, 1987).  We have
operated a 200x200-micron back-illuminated SSPM in a liquid-helium
cooled dewar with a room-temperature transimpedance amplifier output.
Single photon pulses can be easily distinguished above the amplifier
noise.  The individual photon pulses are binned at a selectable time
resolution ranging from $5 \mu$s to 64 ms, and then written to Exabyte
tape.  In the first astronomical application of such a device, we have
made observations of the Crab Nebula pulsar and Her X-1 at
near-infrared wavelengths (J-, H-, and K-bands), and we present the
instrument sensitivities established by these observations.  We
discuss other astronomical observations which are either planned or
currently underway.  Finally, we present design specifications and
predicted performances for a second-generation SSPM high-speed
infrared photometer.

\end{abstract}

\section{Introduction}

	The field of high-speed infrared photometry is one that has
existed for many years, but has until now developed very slowly.  The
range of work done with high-speed near-IR photometers is wide and
varied, from observations of pulsars (both rotation- and
accretion-powered), X-ray binaries, and cataclysmic variables to
applications in IR interferometry, planetary occultations, and lunar
occultations.  However, the amount of work has been limited, largely
due to the relatively poor performance of typical near-IR detectors
(e.g. InSb and HgCdTe - see Lundgren (1994); Pipher {\it et al.},
(1995); Hodapp {\it et al.}, (1995)) at high speed operation (sampling
rates greater than $\sim 100$ Hz).  The appearance of a new
technology, that of the Solid-State Photomultiplier, promises to
change this state of affairs.

	The Solid-State Photomultiplier (SSPM), developed by Rockwell
International Science Center (RISC), is a Si:As detector with
single-photon counting capability over the wavelength range from
$0.4-28$ $\mu$m, intrinsic time response of order 1 ns, and low
detector noise.  As such, it provides great advantages over previous
high-speed IR detectors in terms of both sensitivity and time
resolution.  For instance, our first-generation photometer provides
1000 times the time resolution and 2.6 times the signal-to-noise (at 1
kHz) of the Caltech InSb-based high-speed photometer (Lundgren, 1994).
Thus, the SSPM for the first time brings high-speed IR photometry into
the realm of competition with high-speed optical photometry, which
relies on the well-developed technology of photomultiplier tubes
(PMTs).

\section{Detector}

	The basic operating principle of the SSPM is that of a
blocked-impurity band (BIB) detector operated in avalanche mode.  As
shown in Figure 1(a), a layer of arsenic-doped silicon is deposited on
the silicon substrate, with a blocking layer of undoped Si between the
Si:As layer and the electrode.  A bias applied to the detector results
in the field configuration shown in Figure 1(b).  An incident infrared
photon passes through the substrate and interacts in the Si:As layer,
generating an electron/hole pair.  Under the effect of the electric
field, the electron drifts quickly to the blocking layer.  The
increased field near the blocking layer (BL) (see Figure 1(b))
provides a high cross-section for impact ionization, resulting in an
avalanche of electrons with a typical gain of $\sim 10^4$.  The
electrons are collected at the electrode in timescales of a few
nanoseconds, while the holes drift more slowly to the opposite
electrode implanted in the silicon substrate.  Detector dark current
arises primarily from thermal generation in the Si:As layer.  For a
more detailed description of the SSPM device physics, see Petroff {\it
et al.} (1987).

\section{Cryostat}

	Due to the operating requirements of the SSPM (detector
temperature $< 10 $ K) and its long-wavelength response, the SSPM must
be cooled with liquid helium.  We house the SSPM in an Infrared
Laboratories dewar (see Figure 2 for schematic) with a typical hold
time of $\sim 8-10$ hours.  Light enters the dewar through a fused
silica entrance window, and then passes through a cold filter (usually
a J-, H-, or K-band filter on a thermal blocker which reduces
long-wavelength radiation).  Immediately after the cold filter comes
an aperture stop slide at the focal plane of the telescope, with stop
diameters of 0.6 and 0.3 mm (giving 6 arcsec and 3 arcsec fields on
the Las Campanas Observatory 2.5-meter telescope).  Two fused silica
lenses reimage the aperture stop on the detector with a 3:1 ratio.
The first lens images the telescope secondary mirror onto a Lyot stop
which blocks stray light.  The second lens then brings the light to a
focus on the 200x200 micron back-illuminated SSPM, which is mounted on
a gold-coated plate along with a temperature-sensing diode and a
heating resistor for thermal control.

\section{Optical Alignment}

	In addition to the previously mentioned advantages of speed
and photon-counting capability, the unique structure of the SSPM also
gives it the ability to be easily aligned with an optical system,
which can otherwise be troublesome for back-illuminated single element
detectors.  The basis of the alignment technique is the interface
between the Si:As layer and the undoped Si blocking layer (BL).  At
room temperature, the As acts as a simple dopant, providing a high
concentration of electrons in the conduction band (that is, n-doping).
Thus, the interface between the Si:As and the BL forms a photodiode
junction at room temperature, sensitive to light out to near the
silicon cutoff ($\sim 1.1 \mu$m).  In order for some photons to
penetrate the Si substrate and yet still interact at the junction, we
must use a fairly bright source with a wavelength just below the
cutoff, such as a Nd:YAG $1.06 \mu$m CW laser.

	We performed the alignment of our instrument at the MIT Laser
Spectroscopy Lab, using a 300 mW CW Nd:YAG laser.  The laser beam was
sent into the (open, room-temperature) dewar and aligned with the
smallest aperture stop using an IR viewer, retro-reflecting the beam
to check orthogonality.  We connected the SSPM dewar leads to a simple
circuit such as is commonly used for reversed-bias operation of
photodiodes.  We placed a chopper wheel in the laser beam path, and
connected the SSPM signal to an SR560 amplifier (see below) with both
high- and low-frequency filtering selected to minimize noise while
passing the chopper frequency.  The signal was easily readable on an
oscilloscope, with typical amplitudes of $\sim 100$ mV, and noise
amplitudes of $\sim 10$ mV.  The position of the SSPM plate was then
manually adjusted to provide the maximum signal.

\section{ Electronics}

	Figure 3 show the SSPM electronics chain for telescope
observations.  A variable potentiometer outside the dewar controls the
detector bias, and a $3.9 M \Omega$ resistor acts as a current-limiter
to prevent detector breakdown.  The other terminal of the detector is
connected to a room-temperature transimpedance amplifier (TIA) (R=
$100 M \Omega$) mounted directly outside the dewar.  The TIA acts as a
current-mode preamplifier, converting the avalanche current to an
output voltage.  Photon events produce typical amplitudes of $\sim
120$ mV at the TIA output, while the RMS noise is typically $< 15$ mV.
The TIA is then fed through a Stanford Research SR-560 amplifier with
gain $\sim 5$, which is typically mounted on the telescope near the
instrument.  The amplifier output feeds into a Stanford Research
SR-400 digital photon counter.  We set the SR-400 discriminator level
to cleanly separate real photon events from noise.  One SR-400 output
feeds a PC real-time display, which facilitates target acquisition and
data monitoring throughout the observation.  The second SR-400 output
is a digital logic pulse to a data recording system.  We measure the
time resolution of the entire detector+electronics system (limited by
the post-detector electronics) to be $\sim 0.2 \mu$s, with dead-time
of $\sim 0.8 \mu$s per count.

	The data recording system we use is the "Lil Wizard
Pulsarator" built by Richard Lucinio of Caltech and Jerome Kristian of
the Carnegie Obervatories.  The heart of the Wizard is 2 sets of 2
counting registers.  One register accumulates counts from the SR-400
output, while the other counts a preset number of oscillations from a
Rubidium (Rb) frequency standard (user-selectable register integration
times range from $5 \mu$s to 64 ms).  When the Rb register reaches the
preset limit, the counting function switches to the other set of
registers, and the first data register is transferred to a buffer.
After the buffer accumulates 8192 data elements, it is flushed to an
Exabyte tape drive, along with header information including the number
of Rb oscillations since power-up.  Thus, the Wizard provides
continuous data recording while maintaining phase coherence to the
accuracy of the Rb frequency standard (typical Allan variance $\sim
10^{-12}$ s/s), even over several separate nights of observation.

\section{Laboratory Tests}

	We have evaluated the performance of our SSPM instrument in
the laboratory, using a standard infrared test setup.  A blackbody
cavity with a pinhole aperture illuminates a 45-degree flat mirror
through a variable-speed chopper wheel.  The blackbody beam is either
fed into the SSPM optics directly from the flat mirror (for most
measurements), or else is reflected off a scanning flat mirror, and
then into the SSPM optics (for measuring the instrument beam profile).
We have measured the instrument dark current and JHK quantum
efficiency across a wide range of detector temperatures and bias.
Typical quantum efficiency and dark current are plotted in Figures 4
and 5.  Taking these data, along with estimates or measurements of
background for a band, field of view, and telescope, we can derive the
relative signal-to-noise for the various combinations of temperature
and bias.  Figure 6 shows en example curve for H-band observations on
the MMT with a 2-arcsec field of view.  We use such curves to
select the optimal operating parameters for a given observation.

\section{Observational Results}

	We have used the SSPM high-speed photometer on several
telescopes, including the Whipple Observatory 1.2-meter, the Las
Campanas 2.5-meter, and the Multiple Mirror Telescope.  Some of the
most striking examples of the SSPM's capabilities are shown in Figure
7.  Figure 7(a) shows a J-band ($1.25 \mu$m) pulse profile of the 33
ms Crab Nebula pulsar with $20 \mu$s time resolution.  Taking
advantage of this high time resolution, we have found trends in the
pulse shape as a function of wavelength (Ransom et al., 1994;
Eikenberry et al., 1996ab), including changes in the peak-to-peak
separation and the peak half-widths on timescales $\sim 100 \mu$s,
which are providing new challenges for models of the pulsar emission
mechanism.  Previous infrared photometers have lacked the time
resolution to detect these effects (by an order of magnitude).

	Figure 7(b) shows the H-band pulse profile of Hercules X-1, an
X-ray pulsar.  While the time resolution on these 1.2 s pulses is
unremarkable, the sensitivity of the observation is exceptional.  The
Figure 7(b) profile, which results from a 1 hour observation on the
Whipple 1.2-meter telescope, has a $\sim 4 \sigma$ significance.
Observations with previous detectors (e.g. Middleditch, Pennypacker,
and Burns, 1984) have required 3 hours on 3-meter telescopes to obtain
slightly weaker detections.  From these observations, we have
determined the $3 \sigma$ sensitivity of the instrument to a Crab-like
pulse profile to be $1.4 \times 10^{-4}$ Jy (H=17.3) (1.2-meter
telescope), $3.6 \times 10^{-5}$ Jy (H=18.8) (2.5-meter telescope),
and $1.2 \times 10^{-5}$ Jy (H=20.0) (MMT) for a 1 hour observation.
The SSPM sensitivity scales according to the following equation

\begin{equation}
\nonumber {S\over{N}} = {\eta \ F_{sig} \ t_{obs} \over{\sqrt{[n_D + \eta (F_{sig}+F_{sky}) \ \Delta \lambda \ A \ \Omega] \ t_{obs}}}},
\end{equation}
where S/N is the signal-to-noise ratio, $\eta$ is the SSPM quantum efficiency, $F_{sig}$ is the signal photon flux density (photons per square centimeter per micron per square arcsecond per second), $t_{obs}$ is the observation time in seconds, $n_D$ is the SSPM dark count rate (counts per second), $F_{sky}$ is the sky background photon flux density, $\Delta \lambda$ is the bandpass (microns), $A$ is the telescope area (square centimeters), and $\Omega$ is the field of view (square arcseconds).

	The sensitivities given above also allow a direct comparison
between the SSPM and other types of high-speed infrared photometers.
As none of these others are actual photon counters, their noise will
have a $\sqrt{f}$ dependence, where $f$ is maximum sampling rate - the
electronic bandwidth for analog detectors (e.g. single-pixel devices),
or the read rate for digital detectors (e.g. array detectors).  We
plot the relative signal-to-noise for the SSPM versus a single-pixel
InSb detector (Lundgren, 1994) in Figure 8.  Again, the response curve
for all non-photon-counting detectors will have the shape shown by the
InSb detector, while the SSPM curve is flat with respect to sample
rate.  Thus, at speeds greater than $\sim 100$ Hz, the SSPM becomes
the IR detector of choice.

\section{Other Applications}

	In addition to the observations mentioned above, we are using
the current SSPM photometer in a wide range of other astronomical
applications.  In the area of pulsar research, we are searching for
infrared pulsations from SN1987A and other supernova remnants.  We
have also observed other known pulsars (such as the optical pulsar
0540-69) and compact objects to search for infrared pulsations.  A
brief list of observed targets includes pulsars (PSR0540-69,
PSR1257+12, PSR1957+20, Geminga, PSR1509-58), supernova remnants
(SN1987A), soft gamma-ray repeaters (SGR1806-20, SGR1900+14),
cataclysmic variables (RW Sextans), black hole candidates (J0422+32),
and globular clusters (M15).  While we have found no pulsar signals in
these preliminary searches, we plan to continue these searches in the
future.

	Besides the research geared toward compact objects, the SSPM
may be extremely useful in the field of IR adaptive optics.  In
adaptive optics, one of several corrections is the global wavefront,
or "tip-tilt" correction.  Previously, both global and local wavefront
corrections have been done in the visible, even though the science
detectors usually operate in the near-IR (e.g. Lloyd-Hart {\it et
al.}, 1995).  Since the Earth's atmosphere possesses differential
dispersion between the optical and IR wavebands, the actual distortion
being corrected may differ significantly between the optical and the
IR.  Thus, several groups are investigating the construction of
adaptive optics systems with IR tip-tilt coreection, sampling from
40-100 Hz (M. Lloyd-Hart, personal communication).  While the current
SSPM is competitive with "standard" detectors only at the high end of
this range, work on future SSPM instruments (see next section) will
almost certainly alter this state of affairs dramatically.

\section{Future Instrument Development}

	As mentioned above, the current back-illuminated instrument is
the fastest and most sensitive astronomical high-speed IR photometer
yet built.  However, we have considerable room for improvement,
especially in the area of quantum efficiency (typically $\sim 1 \%$
for J-, H-, and K-bands).  The back-illuminated SSPM is limited in QE
at these wavelengths by the relatively short path length of the
incident photons through the sensitive region (typically $\sim 25-45
\mu$m), compared to the typical absorption length ($\sim 1$ mm).  RISC has found one
solution in the edge-illuminated SSPM, which is similar to the
back-illuminated SSPM, but, as its name suggests, is illuminated from
the side, resulting in a path length of order $1000 \mu$m.  Using this
technique, RISC has achieved near-infrared quantum efficiencies of
$\sim 30-40 \%$ in laboratory tests.  We are currently building a 2nd
generation instrument employing an edge-illuminated SSPM as the
detector.

	The principle problem we have encountered in building this
instrument is the small area$\times$solid-angle product of the
detector.  In order to achieve maximum response, the detector requires
an input beam of f/3 or slower.  However, given the $45 \mu$m
dimension of the sensitive area, this corresponds to a field of view
of only $\sim 0.75$ arcsec on a 4-meter telescope.  In order to
increase this field, we use an optical system that feeds the detector
with an f/1 beam, resulting in a more practical 2.3 arcsec field.  By
calculating the average path length through the detector, we estimate
that this will reduce the effective quantum efficiency by $\sim 35
\%$, to a final value of $\sim 25 \%$.

	Given measurements of sky backgrounds with the present
instrument and of the dark current from the edge-illuminated detector,
we conclude that the edge-illuminated SSPM should produce a factor of
$\sim 6$ improvement in sensitivity over the current SSPM.  Thus, we
expect that our $3 \sigma$ detection limit for a 1 hour observation
will be $2.5 \times 10^{-5}$ Jy (H=19.2) for a 1.2-meter telescope,
$6.3 \times 10^{-6}$ Jy (H=20.7) for a 2.5-meter telescope, and $1.9
\times 10^{-6}$ Jy (H=22.0) for the MMT.

\begin{acknowledgements}

	We thank RISC for supplying the SSPMs, and especially K. Hays
and M. Stapelbroek for their invaluable advice and help, and R.
Florence for his continued support of infrared astronomical instrument
development; J.  Middleditch, C. Pennypacker and J. Kristian for their
support of the SSPM project in both development and observations; J.
Geary, S.  Willner, C. Hughes, and P. Crawford for their assistance in
designing and assembling both instruments at the CfA; D. Paolucci and
the MIT Laser Spectroscopy Lab for help with the laser alignment; Bill
Riley and EG \& G for supplying the Rubidium frequency standard; R.
Narayan for providing funding support; and, R. Lucinio for maintaining
the Wizards.  The SSPM project has been supported in part by a
Smithsonian Scholarly Studies Grant.  S. Eikenberry is supported by a
NASA Graduate Student Researchers Program fellowship through NASA Ames
Research Center.

\end{acknowledgements}

\vfill \eject

\begin{figure}
\vspace*{200mm}
\includegraphics{spfig1.ps}
\caption{(a) SSPM layer configuration with operational schematic, (b) SSPM electric field configuration}
\end{figure}

\begin{figure}
\vspace*{200mm}
\includegraphics{spfig1.ps}
\caption{SSPM instrument schematic layout}
\end{figure}

\begin{figure}
\vspace*{200mm}
\includegraphics{spfig1.ps}
\caption{SSPM electronics layout}
\end{figure}

\begin{figure}
\vspace*{200mm}
\includegraphics{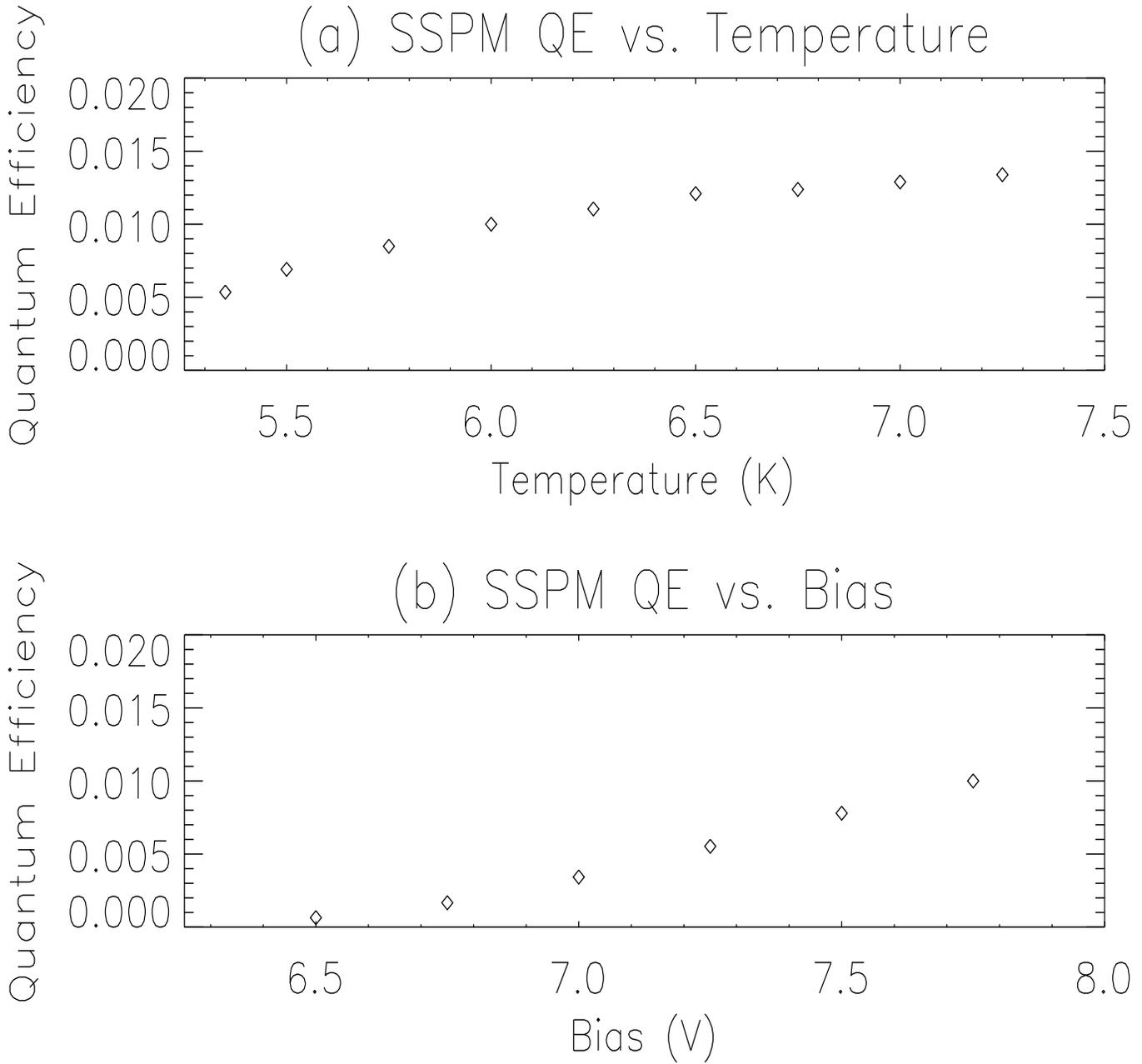}
\caption{(a) SSPM K-band quantum efficiency versus temperature at bias = 7.75 V, (b) SSPM K-band quantum efficiency versus bias at temperature = 6.0 K}
\end{figure}

\begin{figure}
\vspace*{200mm}
\includegraphics{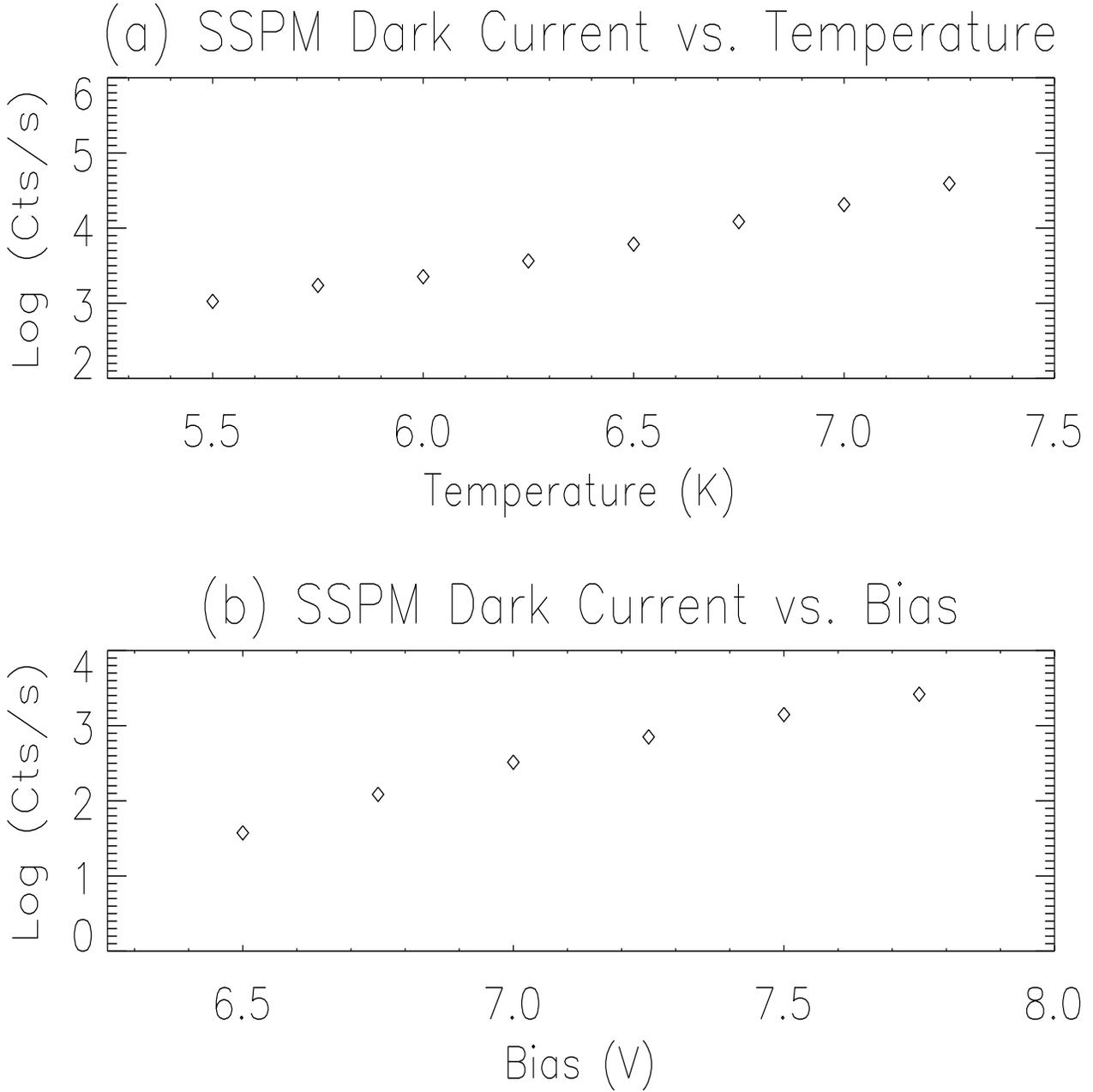}
\caption{(a) SSPM dark current versus temperature at bias = 7.75 V, (b) SSPM dark current versus bias at temperature = 6.0 K}
\end{figure}

\begin{figure}
\vspace*{200mm}
\includegraphics{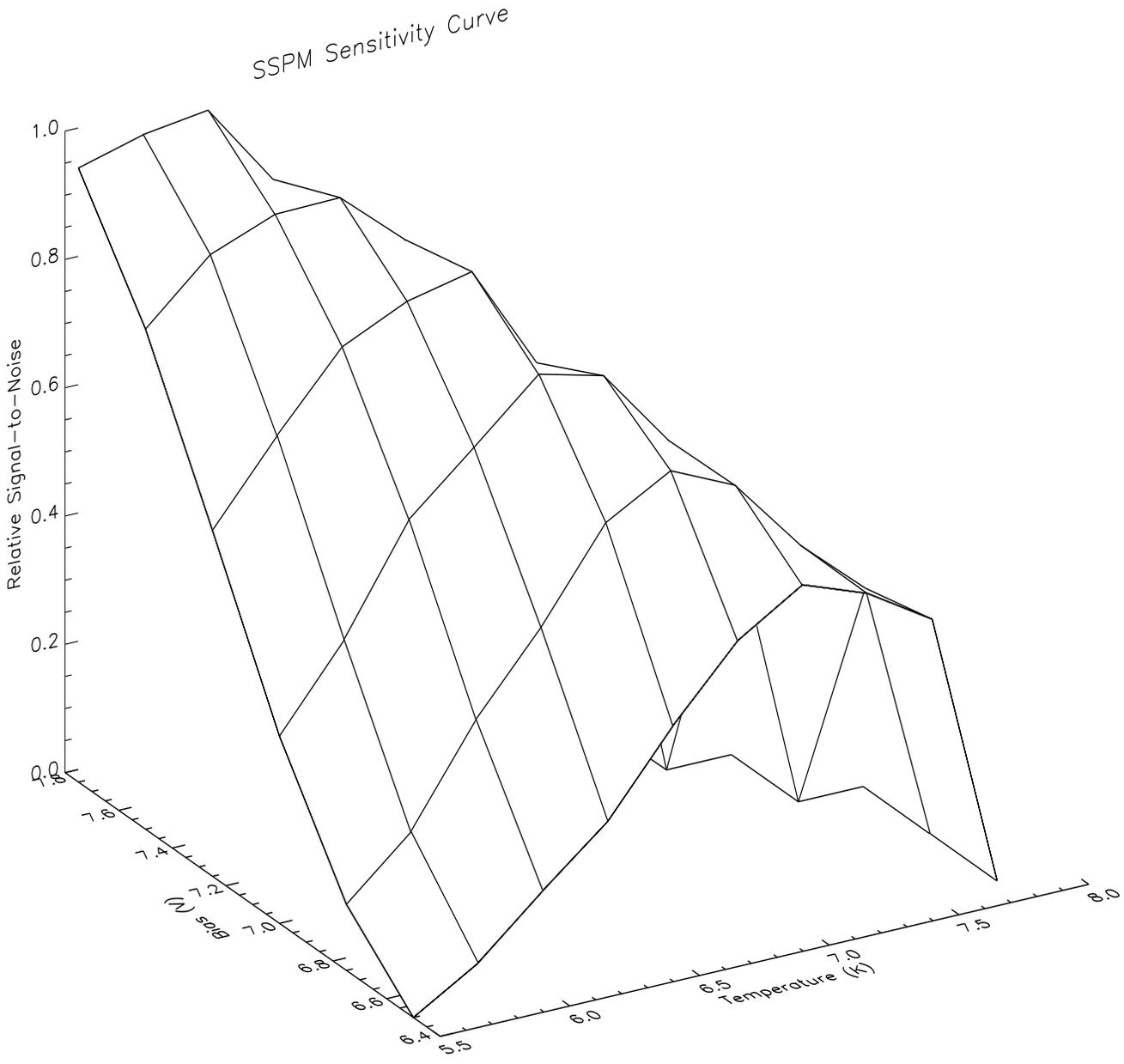}
\caption{SSPM relative signal-to-noise versus temperature and bias}
\end{figure}

\begin{figure}
\vspace*{200mm}
\includegraphics{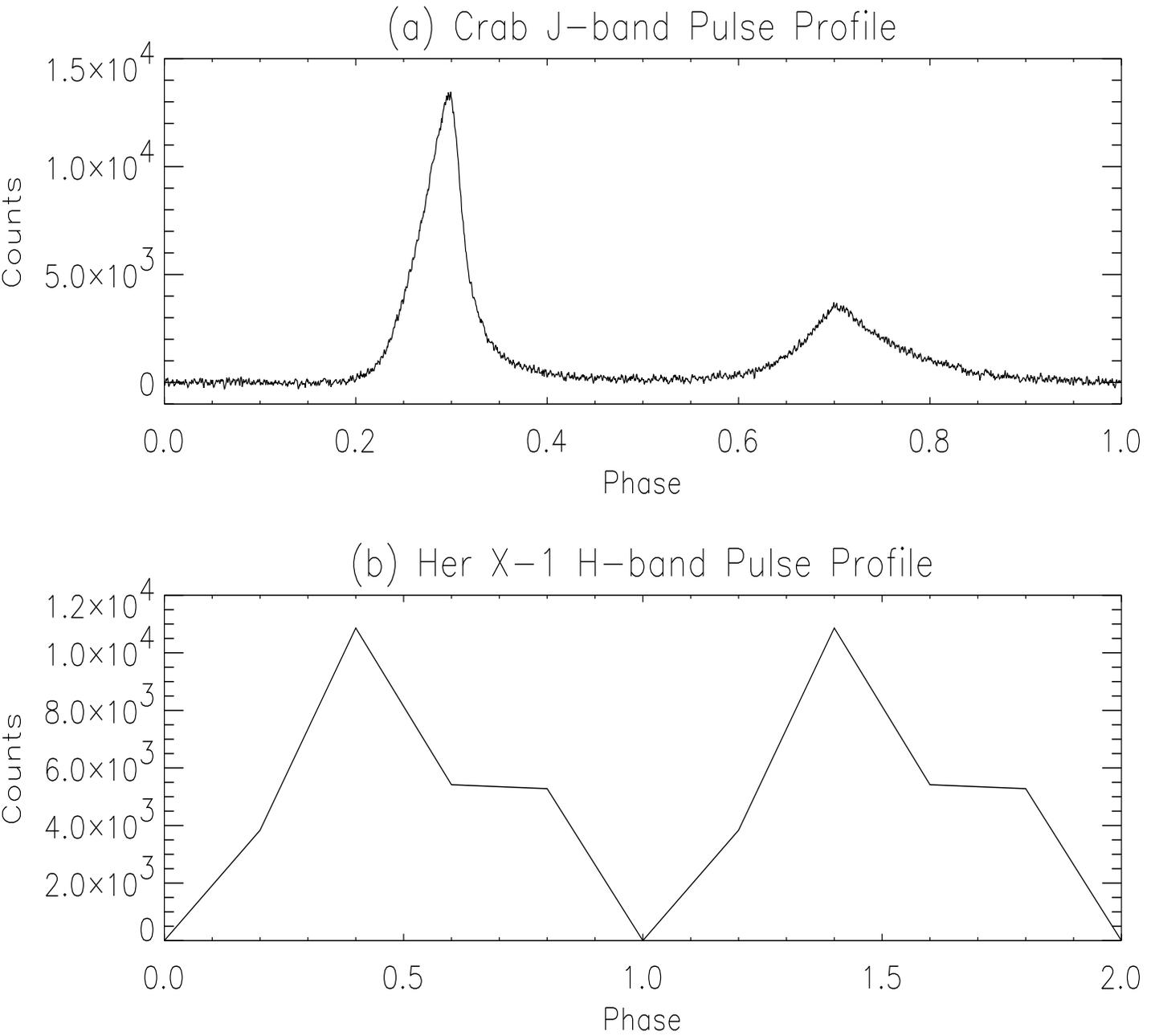}
\caption{(a) Crab Nebula pulsar J-band pulse profile, (b) Her X-1 H-band pulse profile}
\end{figure}

\begin{figure}
\vspace*{200mm}
\includegraphics{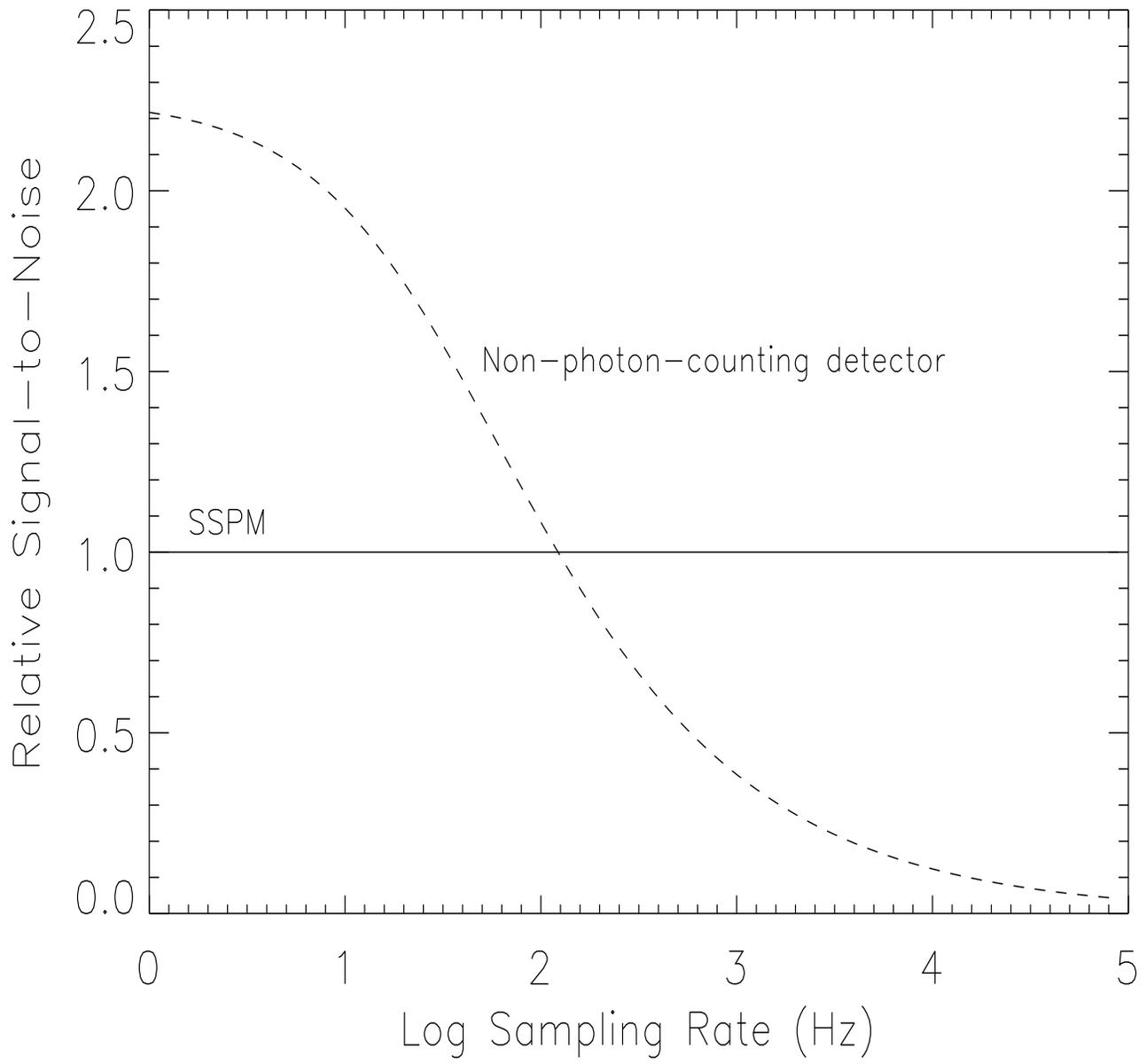}
\caption{Relative signal-to-noise versus sampling rate}
\end{figure}

\end{document}